\documentclass{epl}

\usepackage{amsmath}
\usepackage{amstext}
\usepackage{latexsym}

\newcommand{\ket}[1]{\left\vert#1\right\rangle}

\newlength{\defbaselineskip}
\newcommand{\virgo}[1]{``#1''}
\newcommand{\one}{\mbox{$1 \hspace{-1.0mm}  {\bf l}$ }}  %per la matrice unita

\title{Entanglement production by quantum error correction in the
presence of correlated environment}

\author{G. De Chiara \inst{1}, R. Fazio \inst{1}, C. Macchiavello \inst{2} \and G.M. Palma\inst{3}}
\institute{
  \inst{1} NEST- INFM \& Scuola Normale Superiore, piazza dei Cavalieri 7 , I-56126 Pisa, Italy\\
  \inst{2} INFM \& Dipartimento di Fisica \virgo{A. Volta}, Universit\'a degli Studi di Pavia; Via Bassi 6,I-27100 Pavia, Italy\\
  \inst{3}NEST- INFM \& Dipartimento di Tecnologie dell'Informazione, Universit\'a degli Studi di Milano; via
Bramante 65, I-26013 Crema(CR), Italy }

\pacs{03.67.Pp}{Quantum error correction and other methods for protection against decoherence}
\pacs{03.67.Mn}{Entanglement production, characterization, and manipulation } \pacs{03.65.Yz}{Decoherence; open
systems; quantum statistical methods }

\begin{document}

\maketitle

\date{\today}
\begin{abstract}
We analyze the effect of a quantum error correcting code on the entanglement of encoded logical qubits in the
presence of a dephasing interaction with a correlated environment. Such correlated reservoir introduces
entanglement between physical qubits. We show that for short times the quantum error correction interprets such
entanglement as errors and suppresses it. However for longer time, although quantum error correction is no longer
able to correct errors, it enhances the rate of entanglement production due to the interaction with the
environment.
\end{abstract}

Quantum error correction (QEC) \cite{shor, steane} has been introduced to perform quantum information processing
\cite{QEC} in the presence of noise due to the interaction between quantum bits and environment. The basic idea is
to encode each logical qubit in a redundant way on a set of physical qubits and to periodically acquire
information on the errors that affected the system but not on the quantum state of the system itself. Such
techniques have been developed to deal with independent errors on individual physical qubits due to the
interaction of each physical qubit with its own reservoir. In several physical situations however the presence of
correlated reservoirs \cite{palmasuominen} can result in non-trivial effects. For example it has been shown  that
the interaction of two subsystems with a finite temperature common bath of harmonic oscillators can, for short
times, induce entanglement between the two subsystems initially in a product state. This is possible when the
 environment has some spatial correlations \cite{braun}, as often occurs
 in solid state physics, leading to an
 effective interaction between the two subsystems.
 This effect is also present for noisy baths in the Markovian regime \cite{Flo}.
 The dynamics of the entanglement rate in the presence of decoherence was
also studied \cite{xxyi}.

Quantum error correction has been analyzed in the presence of correlated environments in Ref.\cite{fazioaverin}.
In the present work we address the issue of the effects of QEC on the entanglement between \emph{logical} qubits.
It is reasonable to expect that the entanglement induced by the correlated bath between physical qubits will
modify the encoded state in a way that is interpreted by the QEC procedure as error and therefore corrected.
However, when such entanglement becomes sufficiently large the protocol may be not able to correct it. It is
therefore interesting to study how entanglement is modified by the application of QEC. In the following we will
show that, although QEC is unable to correct such errors, it can enhance the generation of entanglement in a pair
of \emph{logical} qubits with respect to the entanglement induced by the environment on a pair of \emph{physical}
qubits.

The model we consider  in this work, the same as in \cite{palmasuominen}, consists of a register of quantum bits
interacting with a common environment, modelled as a bath of harmonic oscillators. The bath - qubit interaction is
described by the following Hamiltonian

\begin{equation}  \label{hamiltonian}
H = \sum_j \sigma_z^j \xi_j(t)
\end{equation}
where $\xi_j(t)=\sum_{m,\omega} [ \lambda_{j,m}(\omega) a_{m,\omega} +hc]$. In the previous expression
$\lambda_{j,m}(\omega)$ denote the coupling constants between the $j$th qubit and the oscillator at frequency
$\omega$ in the $m$th bath with corresponding annihilation (creation) operator $a_{m,\omega}
(a^\dagger_{m,\omega})$.

In the following we will concentrate our interest on the register dynamics i.e. on the reduced density operator
$\rho(t) = \mathrm{Tr}_E [ U(t) \rho(0) U^\dagger(t)]$ where $\mathrm{Tr}_E$ denotes the partial trace performed
on the environment degrees of freedom. The resulting density matrix can be written in a compact form as
\cite{fazioaverin}

\begin{equation} \label{rot}
\rho(t)=\$ [U_r\rho(0)U_r^\dagger]
\end{equation}
where $\$$ is a map (a super-operator) defined as
\begin{equation}  \label{dollaro}
\$ = \exp\left[ -\frac{1}{2} \sum_{j,j'}\Gamma_{j j'} (\sigma_z^j - \bar\sigma_z^j)(\sigma_z^{j'}
-\bar\sigma_z^{j'})
                \right]\;.
\end{equation}
In the above expression we used the convention that a bar over an operator means that it acts on the density
matrix from the left and $\Gamma_{j j'}(t)=\left\langle \phi_j(t)\phi_{j'}(t)\right\rangle$, with
$\phi_j(t)=\int_0^t \xi_j(t')dt'$, involves correlations of the bath at different times.
 Finally the unitary evolution $U_r$ is given by
\begin{equation} \label{u}
U_r= \exp\left[ i\sum_{j j'} V_{j j'} t\sigma_z^j\sigma_z^{j'}\right]
\end{equation}
where the quantity $V_{jj'}= 2\mathrm{Re} \sum_{m,\omega} \lambda_{m,j}(\omega)\lambda_{m,j'}^*(\omega)/\omega$
 is non zero only if the same reservoir is coupled
to different qubits. As we will see in the following, the unitary operator $U_r$ is responsible for the creation
of entanglement, while the super-operator $\$$  describes the dephasing of the off diagonal elements and is
present also without
 spatial correlations of the environment.

\begin {figure}[ht]
 \centering
  \includegraphics*[scale=0.35]{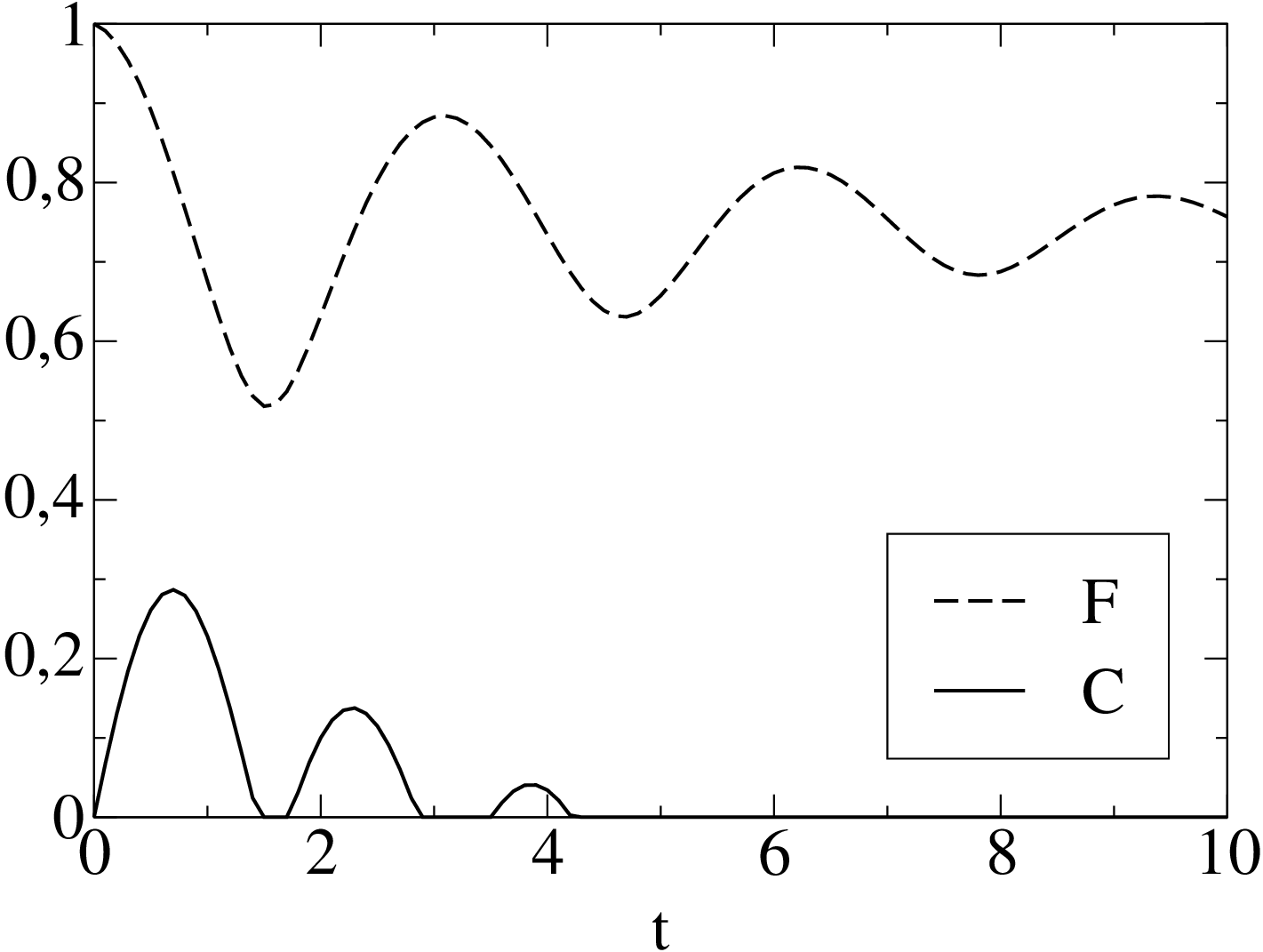}
  \caption {Time evolution of fidelity and concurrence between physical qubits,
averaged over initial states,
  in the absence of  QEC. The time scale is chosen so that $V=1$ and
$\Gamma(t) \simeq 0.1 t $}.
  \label{noqec}
  \end {figure}

As a  measure of entanglement between two qubits we will use the concurrence defined by Wootters \cite{wott}. If
$\rho$ is the density matrix of the global system of two qubits, let us define $\tilde\rho \doteq
\sigma^y\otimes\sigma^y \rho^* \sigma^y\otimes\sigma^y$ and $R=\rho\tilde\rho$. The concurrence is then defined as
$ C = \max \{0, \lambda_1-\lambda_2-\lambda_3-\lambda_4 \}$ where $\lambda_i$ are the square roots of the
eigenvalues of $R$ labelled in decreasing order. Using this definition we obtain for Bell states $\ket{\phi^\pm}
=(\ket{00}\pm\ket{11})/\sqrt{2}$ the result $C(t) = e^{-4(\Gamma_{11}+2\Gamma_{12}+\Gamma_{22})}$ while for
$\ket{\psi^\pm}=(\ket{01}\pm\ket{10})/\sqrt{2}$ the concurrence decays as $C(t) =
e^{-4(\Gamma_{11}-2\Gamma_{12}+\Gamma_{22})}$. It is easy to see that if the environment has special symmetries,
for instance if $\Gamma_{ij}$ is equal for different $i$ and $j$  then  the subspace spanned by $\ket{\psi^\pm}$
is decoherence free \cite{palmasuominen,zanardi}. Note  that the typical decoherence times are of the order of
$dt/d\Gamma_i$. If we choose $\ket{ij}_x$, eigenstates of $\sigma_x$ as initial states, it is possible to see that
the qubits become entangled   if $V_{12} > d\Gamma_i/dt$. In fact the entanglement oscillates with a frequency
$4V_{12}$  and is damped with a rate proportional to $d\Gamma_i/dt$ (see figure \ref{noqec}).
%%%%%%%%%%%%%%%%%%%%%%%%%%%%%%%%%%%%%%%%%%%%%
%%%%%%%%%%%%%%%%%%%%%%%%%%%%%%%%%%%%%%%%%%%%%%%

Let us now introduce a QEC protocol. In the following we will
encode each logical qubit into three physical qubits as follows
$\ket{\tilde 0} \to \ket{000}_x$ and $\ket{\tilde 1} \to
\ket{111}_x$ where $\ket{0}_x$ and $\ket{1}_x$ are eigenstates of
$\sigma_x$. This code will protect the  logical qubits against one
single phase error on individual qubits \cite{shor}. When
$VT,\Gamma(T) \ll 1$, where $T$ is the time interval between two
quantum error corrections, following the same approach of
\cite{fazioaverin} we can write the following master equation for
the continuous evolution of the density matrix of the two logical
qubits

\begin{equation} \label{me2}
\frac{d\rho}{dt} = -(\gamma_1 + \gamma_2)\rho + \gamma_1 \sigma_x^1 \rho \sigma_x^1 + \gamma_2 \sigma_x^2 \rho
\sigma_x^2
\end{equation}

where

\begin{equation}
\gamma_i = \frac{2}{T} \sum_{j>j'} (T^2V^2_{jj'} + 2\Gamma_{jj'}^2(T)+ \Gamma_{jj}(T)\Gamma_{j'j'}(T))\;.
\end{equation}
as in Ref.\cite{fazioaverin}. In the above equation we have to take $j=1,2,3$ for $i=1$ and $j=4,5,6$ for $i=2$.
The rates $\gamma_i\ll d\Gamma_{ij}/dt$ are the same found in \cite{fazioaverin} for the single qubit case.

Let us now turn our attention to the time evolution of the
entanglement between logical qubits in the presence of QEC in the
regime in which the above master equation holds. As a first
example we consider the initial state $\ket{\tilde 0\tilde 0}$. In
this case the only non-vanishing elements at time $t$ are those in
the main diagonal, and they decay exponentially with decay
constants equal to $\gamma_1 ,\gamma_2$ and $\gamma_1+\gamma_2$.
If instead we start with a Bell state $\tilde{\ket{\phi^+}}$ of
logical qubits the non-zero elements at time $t$ are those on the
two diagonals. Such state evolves towards an incoherent
superposition of two Bell states whose concurrence decays as
$C(t)=e^{-2(\gamma_1+\gamma_2)t}$ and the fidelity with respect to
the initial state is
$F(t)=\frac{1}{2}(1+e^{-2(\gamma_1+\gamma_2)t})$, where, for an
initially pure state $|\psi (t=0)\rangle$ the fidelity is defined
as $F(t) = \langle \psi (0) |\rho (t)|\psi (0)\rangle $. From
these results it is evident that the fidelity and the concurrence,
in the presence of QEC, decay with the same rate. It is
interesting also to see what happens to the state $\ket{\tilde i
\tilde j}_x$. In this case the density operator does not evolve:
$\rho(t)=\rho(0)$. This does not follow from the decoherence free
subspace phenomenon but rather from the fact that QEC freezes the
unitary evolution by means of repeated measurement, as in the
quantum Zeno effect \cite{misra}, and conditional dynamics
depending on the measurement outcome.

In order  to obtain information about the global properties of the map we will consider states with fixed initial
concurrence, which are of the form:
\begin{equation}
\ket{\Psi}=\cos\vartheta\ket{\tilde0_{n1}\tilde0_{n2}}+\sin\vartheta\ket{\tilde1_{n1}\tilde1_{n2}}
\label{thetaconc}
\end{equation}
with
\begin{eqnarray}
\ket{\tilde0_{ni}}&=&\cos\frac{\vartheta_i}{2}\ket{\tilde0}+\sin\frac{\vartheta_i}{2}e^{i\varphi_i}\ket{\tilde1}\\
\ket{\tilde1_{ni}}&=&\cos\frac{\vartheta_i}{2}\ket{\tilde1}- \sin\frac{\vartheta_i}{2}e^{-i\varphi_i}\ket{\tilde0}
\end{eqnarray}
The concurrence of $\ket{\Psi}$ is $\sin2\vartheta $. In the following analysis we will average over
$\vartheta_1,\vartheta_2,\varphi_1,\varphi_2$. We can see that averaging over all initial product states
$\vartheta = 0 , \pi/2$ one finds that the concurrence is always zero and the fidelity decays with a rate
proportional to $\gamma_i$. For partially or maximally entangled states the entanglement decays monotonically
again with decay constants $\gamma_1+\gamma_2$. This means that QEC suppresses the effective interaction between
logical qubits due to the presence of a correlated environment. In other words the bath does not induce
entanglement between logical qubits and the entanglement initially present decays. This implies that for small $T$
there is not creation of entanglement since the QEC protocol destroys all the correlations between physical qubits
of different logical qubits. The reason for this can be seen in a qualitative way: for a state $\ket{\tilde
0\tilde 0} = \ket{000}_x\ket{000}_x$ the operator $U_r = \one + i\sum_{ij} V_{ij}t \sigma_z^i\sigma_z^j$ up to
first order creates superpositions like $\ket{000}_x\ket{000}_x + \ket{110}_x\ket{000}_x + \cdots$ which is
entangled. After the corrections one gets  a mixture of $\ket{000}_x\ket{000}_x$ and $\ket{111}_x\ket{000}_x$
which is no longer entangled. Not surprisingly then QEC inhibits the  production of entanglement as this is seen
by the protocol as an error.

\begin{figure}[htbp]
\begin{center}
\includegraphics*[scale=0.35]{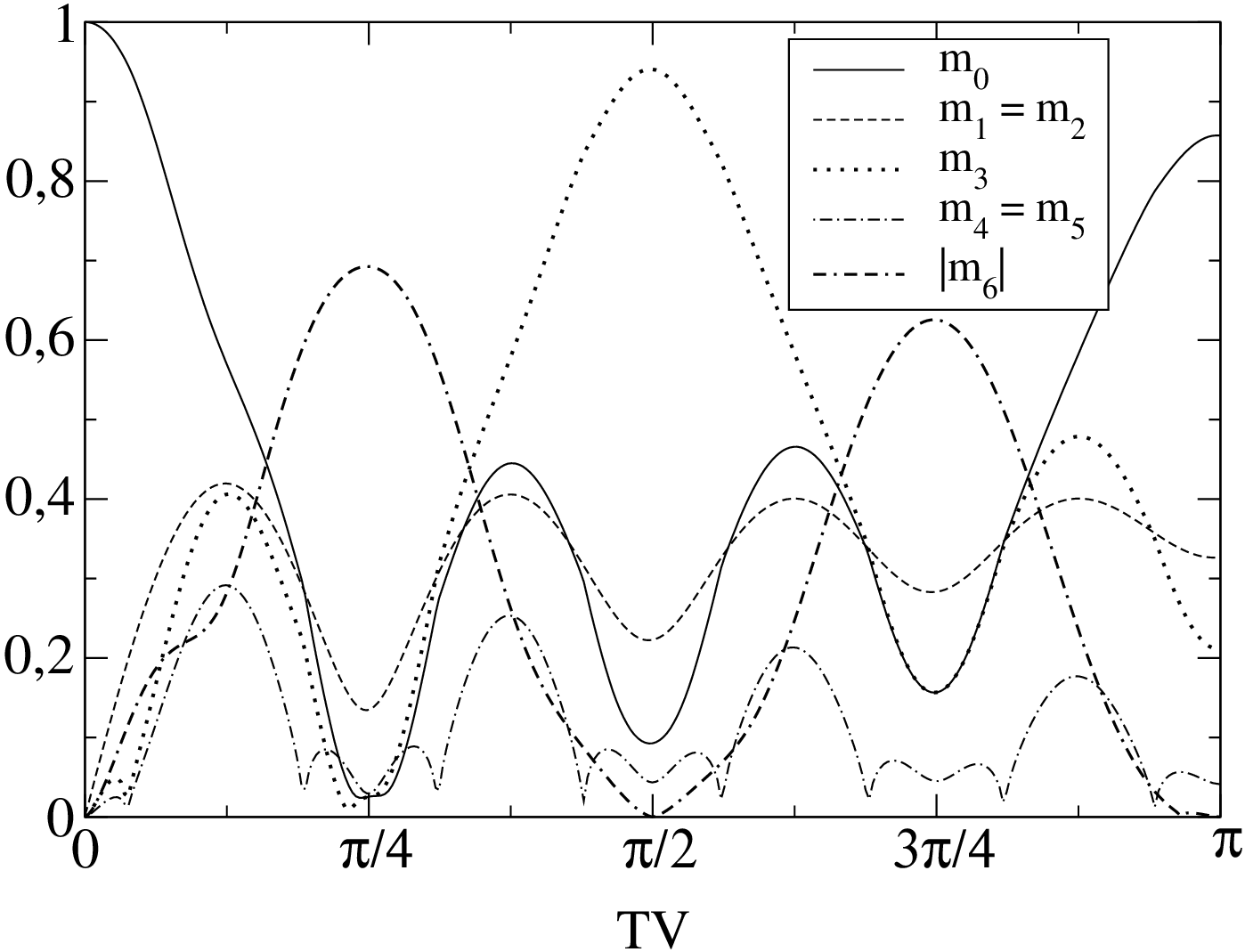}
\caption{Relative weights of the Kraus operators, i.e. the quantities $m_i$, as  functions of $T$. For simplicity
we assume $\Gamma_{ij} (T) =\Gamma(T) \simeq 0.1 T$. From this it follows that $m_1=m_2$.} \label{figurekraus}
\end{center}
\end{figure}

\begin{figure}[htbp]
\begin{center}
\includegraphics*[scale=0.35]{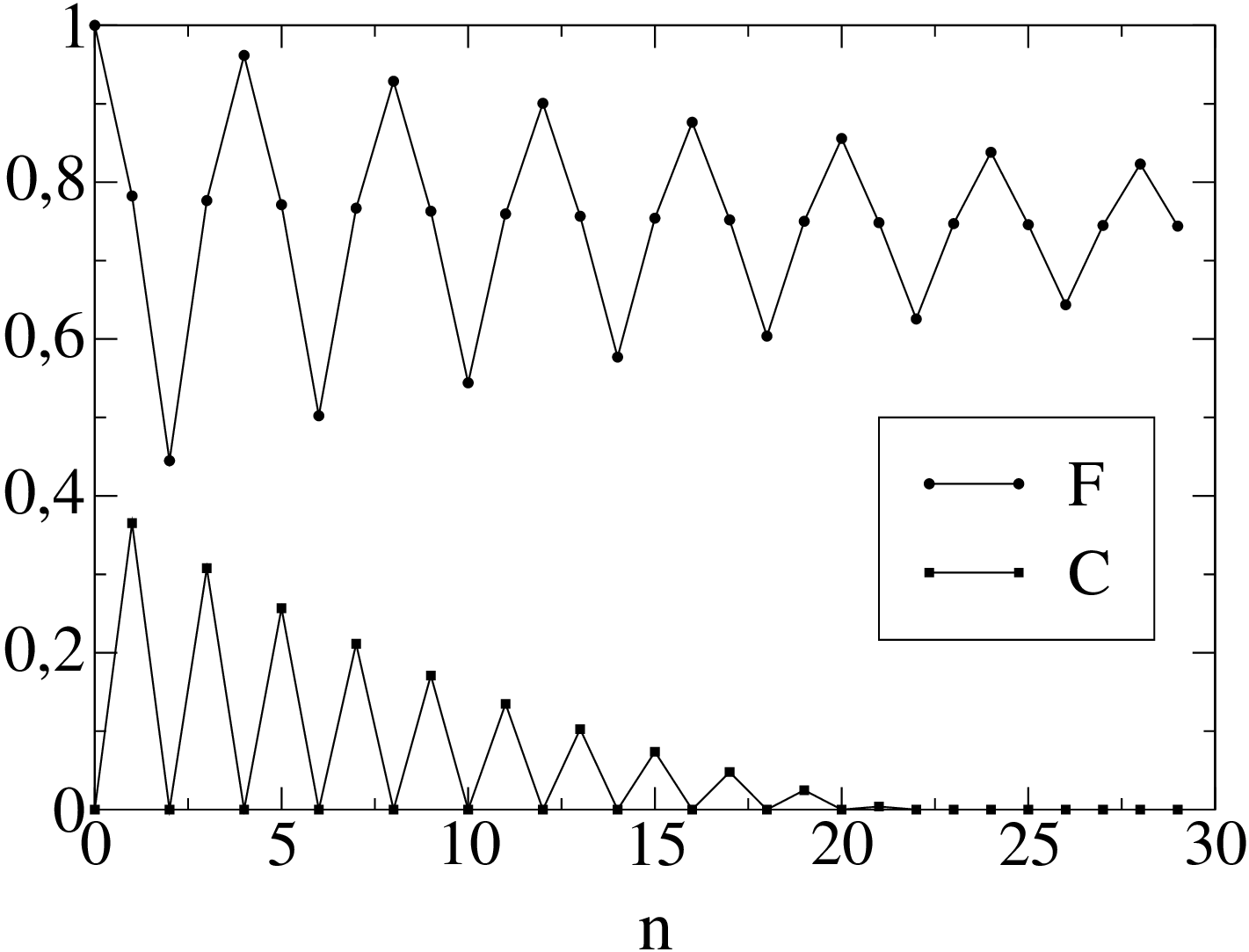}
\caption{Mean fidelity and concurrence between logical qubits in the presence of QEC as functions of the number of
applications of the map, for $T=\pi/4V$, $V=1$ and $\Gamma(t) \simeq 0.1 t $, the same parameters used in figure
\ref{noqec}.} \label{tpi4}
\end{center}
\end{figure}

This may no longer be true if the time $T$ is comparable to the period of oscillation of  entanglement without
QEC. In this case the environment may have time to create enough entanglement to be interpreted as a property of
the initial state and amplified by QEC. If the time $T$ between two corrections is not short the approximations
which lead to the master equation \eqref{me2} are no longer valid and we must use the exact map which links
$\rho(t)$ and $\rho(t+T) $ after the free time evolution and the QEC. In terms of Kraus operators such map can be
written as
\begin{equation} \label{kraus}
\rho_C(t+T) = \sum_k M_k \rho(t) M_k^\dagger \
\end{equation}
where $\sum_k M_k^\dagger  M_k =\one$. In our case we found the following seven Kraus operators:
\begin{eqnarray}
M_0 &=& m_0 \one \notag\\
M_1 &=& m_1 \sigma_{x1}\notag\\
M_2 &=& m_2 \sigma_{x2}\notag\\
M_3 &=& m_3 \sigma_{x1}\sigma_{x2}\notag\\
M_4 &=& m_4 (\sigma_{x1}+\sigma_{x2})\notag\\
M_5 &=& m_5(\one+\sigma_{x1}\sigma_{x2})\notag\\
M_6 &=& |m_6|(\one \pm i\sigma_{x1}\sigma_{x2})
\end{eqnarray}
Such Kraus operators are linked to the $\Gamma_{ij} (T)$ rates,
however the expressions which explicit such dependence are lengthy
and not of immediate reading. A more straightforward physical
picture of the action of the map is instead gained by considering
the plot of the coefficients $m_i$
 as functions of $T$ (see Fig.\ref{figurekraus}). We supposed that
$\Gamma_{ij} (T) =\Gamma(T) \simeq 0.1 T$ from which it follows
that $m_1=m_2$. Notice that in general $m_4 = m_5$ and that the
sign in $M_6$ is plus or minus depending on the sign of
$\mathrm{Im} ( m_6)$.
%%%%%%%%%%%%%%%%%%%%%%%%%%%%%%%%%%%%%%%%%%%
%%%%%%%%%%%%%%%%%%%%%%%%%%%%%%%%%%%%%%%%%%%%
Figure \ref{figurekraus} shows clearly various regimes for the evolution of the logical qubits. Indeed for $VT\ll
1$ only $m_0$ and $m_1$ are non zero, leading to the dissipative master equation \eqref{me2}. As we have seen in
this regime there is no creation of entanglement because $M_{1,2}$ are single qubit operators. When $T$ is close
to the value $\pi/8V$ there are also contributions from the two-qubit operators $M_{4,5,6}$ but the dissipative
effect of $M_{1,2,3}$ is predominant and again we verified that the entanglement production is zero. The
interesting regime is for $T\simeq \frac{\pi}{4V}$, when the main contribution comes from $M_6$. As expected in
this case there is creation of entanglement that oscillates between zero and a maximum value (which will gradually
decay). Indeed for the particular choice  $T = \frac{\pi}{4V}$ (see figure \ref{tpi4})  the created entanglement
is maximum and the QEC amplifies it. The other non zero dissipative contribution is $M_1$ and it is present only
if  $\Gamma_i$ is non zero. On the other hand, if $\Gamma_i$ is negligible, $M_6$ is the only  Kraus operator, and
therefore the evolution is unitary even in the presence of QEC. Under this condition this evolution induces the
transformation $\ket{\tilde 0} \to \ket{\tilde 1}$ and vice versa. For larger $T$ one again finds  regimes with no
entanglement production.
%%%%%%%%%%%%%%%%%%%%%%%%%%%%%%%%%%%%%%%%%%%
%%%%%%%%%%%%%%%%%%%%%%%%%%%%%%%%%%%%%%%%%%%%

When the map (\ref{kraus}) is applied (for $\Gamma_i=0$ and $T =
\frac{\pi}{4V}$)  to the state $\ket{\tilde 0\tilde 0}$ then one
sees that it makes a sort of periodic oscillation:
\begin{equation}
\ket{\tilde 0\tilde 0} \rightarrow \ket{\tilde0 \tilde0}+i\ket{\tilde1 \tilde1} \rightarrow \ket{\tilde 1 \tilde
1}
 \rightarrow \ket{ \tilde 0 \tilde 0}-i\ket{\tilde1 \tilde1} \rightarrow \ket{\tilde0\tilde0}
\end{equation}
 Of course if $\Gamma$ is not zero then the other Kraus operators also
contribute and so this oscillation is damped. This type of oscillation is not restricted to initial product states
but also to partially entangled states. As a consequence it is possible to create entanglement that can be
distilled to obtain maximally entangled states or even extracted from the system.

Figure \ref{tpi4} shows that the maximum average concurrence
achievable is around 0.4. It is important to underline that this
is an average value: there are states, like $\ket{\tilde 0\tilde
0}_x$, that do not evolve and so there is no production of
entanglement while there are states for which the created
entanglement is more than the average value. This is the case of
initial product states of the computational basis for which the
production of entanglement is maximum. For example, let us compare
the initial states that lead to the maximum production of
entanglement with and without QEC. In the presence of QEC the best
case is $\ket{\tilde0 \tilde0}$: the entanglement is $0.92$ after
just one application of QEC. On the other hand the maximum
entanglement reached in the case without QEC is only $0.74$ for
the state $\ket{00}_x$. These results are insensitive to the
change $0 \to 1$.

In summary, we studied the effect of QEC on the entanglement between logical qubits in the presence of correlated
noise. We found that when the time interval $T$ between consecutive applications of QEC is small QEC is helpful in
preserving the state of the qubits and indeed entanglement decays with a smaller rate. As expected, in the absence
of QEC there is production of entanglement between physical qubits, due to the correlations introduced by the
environment, while in the presence of QEC this phenomenon is inhibited. The interesting new effect we have
discovered is that when $T$ is not small then entanglement may be generated with a bigger rate than in absence of
QEC. In this scenario we are in a situation in which the entanglement induced by the unitary dynamics generated by
the correlated bath is enhanced by the QEC protocol which prevents the dephasing effects of the coupling with the
bath. Note that such enhancement in the production rate of entanglement is achieved by means of local measurements
and conditional local unitary operations on the logical qubits, in other words the entanglement is not induced by
joint measurements on the pair of logical qubits.

Our results suggest a new strategy to enhance the rate of entanglement production for interacting qubits in the
presence of decoherence in a more general scenario. Consider a system of interacting subsystems in the presence of
a decohering dynamics. Such situation might be due either to a coupling of the subsystem to a common reservoir,
like in our case considered above, or to a direct mutual coupling between subsystems in the presence of
independent environments. The entanglement induced between subsystems by the unitary dynamics is destroyed by the
dissipative one. By enlarging the Hilbert space, in our case by increasing the number of qubits, it is possible to
identify separate larger subsystems such that by means of local measurements and local conditional unitary
operations on such subsystems the amount of entanglement due to the mutual interaction is increased with respect
to the entanglement it would have been generated between the original subsystems. In other words we conjecture
that the strategy outlined above is not restricted to QEC but it is a more general technique. The issue is,  given
an effective interaction in the presence of decoherence, to find optimal subsystems and projection protocols in
order to maximize the production of entanglement.

In some sense our protocol can be described as a generalization of the one proposed in  Ref. \cite{dur}. In
\cite{dur} the entanglement production is optimized, in the presence of a direct interaction and in the absence of
decoherence, by means of local operations and ancillas. In our QEC protocol we use some sort of ancillary system
to enlarge the Hilbert space, although in this case there is not a sharp distinction between qubits and ancillas.
Furthermore, we make use also of local projections on the enlarged subsystems and conditional dynamics.

We gratefully acknowledge many helpful discussions with D. Averin. This work was supported by the EU (IST-SQUBIT,
HPRN-CT-2002-00144; QUPRODIS: IST-2002-38877).

\end{document}